\documentclass[12pt,fleqn,a4paper]{article} 
\usepackage{epsfig,graphics,graphicx}
\usepackage{clshan-math,clshan-dm-dd}
\usepackage{multirow}
\def \lsim {\:\raisebox{-0.7ex}{$\stackrel{\textstyle<}{\sim}$}\:}
\def \gsim {\:\raisebox{-0.7ex}{$\stackrel{\textstyle>}{\sim}$}\:}
\begin{document}
\thispagestyle{empty}
\begin{flushright}
 March 2015
\end{flushright}
\begin{center}
{\Large\bf
 Reconstructing the WIMP Velocity Distribution            \\
 from Direct Dark Matter Detection Data                   \\ \vspace{0.2 cm}
 with a Non--Negligible Threshold Energy}                 \\
\vspace*{0.7cm}
 {\sc Chung-Lin Shan} \\
\vspace*{0.5cm}
 {\it Xinjiang Astronomical Observatory,
      Chinese Academy of Sciences                          \\
      No.~150, Science 1-Street,
      \"{U}r\"{u}mqi, Xinjiang 830011, China}              \\~\\
\vspace{0.05cm}
 {\it E-mail:} {\tt clshan@xao.ac.cn}                      \\
\end{center}
\vspace{1cm}
\begin{abstract}
 In this paper,
 we investigate the modification of our expressions
 developed for
 the model--independent data analysis procedure of
 the reconstruction of the (time--averaged) one--dimensional
 velocity distribution of Galactic
 Weakly Interacting Massive Particles (WIMPs)
 with a non--negligible experimental threshold energy.
 Our numerical simulations
 show that,
 for a minimal reconstructable velocity of
 as high as \mbox{${\cal O}(200)$ km/s},
 our model--independent modification of
 the estimator for the normalization constant
 could provide
 precise reconstructed velocity distribution points
 to match the true WIMP velocity distribution
 with a \mbox{$\lsim$ 10\%} bias.
\end{abstract}
\clearpage
\section{Introduction}

 Currently,
 direct Dark Matter detection experiments
 searching for Weakly Interacting Massive Particles (WIMPs)
 are one of the promising methods
 for understanding the nature of Dark Matter (DM)
 and identifying them among new particles produced at colliders
 as well as
 studying the (sub)structure of our Galactic halo
 \cite{SUSYDM96,
       Drees12,
       Strigari12b,
       Baudis12c}.

 In our earlier work
 \cite{DMDDf1v},
 we developed model--independent methods
 for reconstructing the (moments of the)
 time--averaged one--dimensional velocity distribution of halo WIMPs
 by using
 the measured recoil energies directly.
 However,
 with a few hundreds or even thousands recorded WIMP events,
 only estimates of the reconstructed velocity distribution
 with pretty large statistical uncertainties
 at a few ($<$ 10) points
 could be obtained.
 Hence,
 in order to provide more detailed information
 about the WIMP velocity distribution,
 we introduced the Bayesian analysis
 into our
 reconstruction procedure
 for concretely determining,
 e.g.~the position of the peak of
 the one--dimensional velocity distribution function
 and the values of
 the characteristic Solar and Earth's Galactic velocities
 \cite{DMDDf1v-Bayesian}.

 In our Monte Carlo simulations
 presented in Refs.~\cite{DMDDf1v, DMDDf1v-Bayesian},
 the minimal experimental cut--off energies
 of data sets to be analyzed
 are assumed to be negligible.
 For experiments with heavy target nuclei,
 e.g.~Ge or Xe,
 and once WIMPs are heavy
 (\mbox{$\gsim~100$ GeV}),
 the systematic bias caused by this assumption
 should be acceptable.
 However,
 once WIMPs are light
 (\mbox{$\lsim~50$ GeV})
 and a light target nucleus,
 e.g.~Si or Ar,
 is used
 for reconstructing the WIMP velocity distribution $f_1(v)$,
 effects of a non--negligible threshold energy
 has to be considered and
 the estimate of the normalization constant of $f_1(v)$
 would need to be modified properly.
 Therefore,
 as a supplement of our earlier works,
 we consider in this paper
 the needed modification of
 the normalization constant of $f_1(v)$
 for the general case
 with a non--negligible experimental threshold energy.

 The remainder of this paper is organized as follows.
 In Sec.~2,
 we first review the model--independent method
 for reconstructing the time--averaged
 one--dimensional velocity distribution of halo WIMPs
 by using data from direct DM detection experiments directly.
 Then,
 in Sec.~3,
 we develop
 the modification
 of the normalization constant of $f_1(v)$
 for a non--zero minimal experimental cut--off energy
 step by step.
 Numerical results of
 the modified reconstruction of
 the WIMP velocity distribution
 based on the Monte Carlo simulation
 will be
 given.
 A systematic bias
 in this model--independent modification
 will be discussed particularly.
 We conclude in Sec.~4.
 Some technical details for our analysis
 will be given in Appendix.

\section{Model--independent reconstruction of
         the one--dimensional WIMP velocity distribution}

 In this section,
 we first review in brief the model--independent method
 for reconstructing the (time--averaged)
 one--dimensional
 WIMP velocity distribution
 by using experimental data,
 i.e.~measured recoil energies,
 directly from direct detection experiments.
 Detailed derivations and discussions
 can be found in Ref.~\cite{DMDDf1v}.

\subsection{From the recoil spectrum}
 The basic expression for the differential event rate
 for elastic WIMP--nucleus scattering is given by \cite{SUSYDM96}:
\beq
   \dRdQ
 = \calA \FQ \int_{\vmin}^{\vmax} \bfrac{f_1(v)}{v} dv
\~.
\label{eqn:dRdQ}
\eeq
 Here $R$ is the direct detection event rate,
 i.e.~the number of events
 per unit time and unit mass of detector material,
 $Q$ is the energy deposited in the detector,
 $F(Q)$ is the elastic nuclear form factor,
 $f_1(v)$ is the one--dimensional velocity distribution function
 of the WIMPs impinging on the detector,
 $v$ is the absolute value of the WIMP velocity
 in the laboratory frame.
 The constant coefficient $\calA$ is defined as
\(
        \calA
 \equiv \rho_0 \sigma_0 / 2 \mchi \mrN^2
\),
 where $\rho_0$ is the WIMP density near the Earth
 and $\sigma_0$ is the total cross section
 ignoring the form factor suppression.
 The reduced mass $\mrN$ is defined by
\(
        \mrN
 \equiv \mchi \mN / \abrac{\mchi + \mN}
\),
 where $\mchi$ is the WIMP mass and
 $\mN$ that of the target nucleus.
 Finally,
 $\vmin$ is the minimal incoming velocity of incident WIMPs
 that can deposit the energy $Q$ in the detector:
\(
   \vmin
 = \alpha \sqrt{Q}
\)
 with the transformation constant
\beq
        \alpha
 \equiv \sfrac{\mN}{2 \mrN^2}
\~,
\label{eqn:alpha}
\eeq
 and $\vmax$ is the maximal WIMP velocity
 in the Earth's reference frame,
 which is related to
 the escape velocity from our Galaxy
 at the position of the Solar system,
 $\vesc$.

 In our earlier work \cite{DMDDf1v},
 it was found that,
 by using a time--averaged recoil spectrum $dR / dQ$
 and assuming that no directional information exists,
 the normalized one--dimensional velocity distribution function
 of incident WIMPs
 can be solved
 from Eq.~(\ref{eqn:dRdQ}) directly as
\beq
   f_1(v)
 = \calN
   \cbrac{ -2 Q \cdot \dd{Q} \bbrac{ \frac{1}{\FQ} \aDd{R}{Q} } }\Qva
\~,
\label{eqn:f1v_dRdQ}
\eeq 
 where the normalization condition:
\beq
    \intz f_1(v) \~ dv
 =  1
\label{eqn:normalization_infty}
\eeq
 has been used and thus
 the normalization constant $\calN$ is given by
\beq
   \calN
 = \frac{2}{\alpha}
   \cbrac{\intz \frac{1}{\sqrt{Q}}
                \bbrac{ \frac{1}{\FQ} \aDd{R}{Q} } dQ}^{-1}
\~.
\label{eqn:calN_int}
\eeq
 Here the integral
 goes over the entire physically allowed range of recoil energies:
 starting at \mbox{$Q = 0$},
 and the upper limit of the integral has been written as $\infty$.
 Note that,
 the velocity distribution function of halo WIMPs
 reconstructed by Eq.~(\ref{eqn:f1v_dRdQ})
 is independent of the local WIMP density $\rho_0$
 as well as
 of the WIMP--nucleus cross section $\sigma_0$.
 However,
 not only the overall normalization constant $\calN$
 given in Eq.~(\ref{eqn:calN_int}),
 but also the shape of the velocity distribution
 transformed
 through
 $Q = v^2 / \alpha^2$
 depends on the WIMP mass $\mchi$
 (involved in the coefficient $\alpha$
  defined in Eq.~(\ref{eqn:alpha})).

\subsection{From experimental data directly}
 In order to avoid some model dependence
 during giving a functional form for the recoil spectrum $dR / dQ$
 needed in Eqs.~(\ref{eqn:f1v_dRdQ}) and (\ref{eqn:calN_int}),
 expressions that allow to reconstruct $f_1(v)$
 directly from data
 (i.e.~measured recoil energies)
 have also been developed \cite{DMDDf1v}.

 Consider
 experimental data described by
\beq
     {\T Q_n - \frac{b_n}{2}}
 \le \Qni
 \le {\T Q_n + \frac{b_n}{2}}
\~,
     ~~~~~~~~~~~~ 
     i
 =   1,~2,~\cdots,~N_n,~
     n
 =   1,~2,~\cdots,~B.
\label{eqn:Qni}
\eeq
 Here the entire experimental possible
 energy range between the minimal and maximal cut--offs
 $\Qmin$ and $\Qmax$
 has been divided into $B$ bins
 with central points $Q_n$ and widths $b_n$.
 In each bin,
 $N_n$ events will be recorded.
 Since the recoil spectrum $dR / dQ$ is expected
 to be approximately exponential,
 in order to approximate the spectrum
 in a rather wider range,
 instead of the conventional standard linear approximation,
 the following exponential ansatz
 for the measured recoil spectrum
 (before normalized by the exposure $\calE$)
 in the $n$th bin has been introduced \cite{DMDDf1v}:
\beq
        \adRdQ_{{\rm expt}, \~ n}
 \equiv \adRdQ_{{\rm expt}, \~ Q \simeq Q_n}
 \equiv r_n  \~ e^{k_n (Q - Q_{s, n})}
\~.
\label{eqn:dRdQn}
\eeq
 Here $r_n = N_n / b_n$ is the standard estimator
 for $(dR / dQ)_{\rm expt}$ at $Q = Q_n$,
 $k_n$ is the logarithmic slope of
 the recoil spectrum in the $n$th $Q-$bin,
 which can be computed numerically
 from the average value of the measured recoil energies
 in this bin:
\beq
   \bQn
 = \afrac{b_n}{2} \coth\afrac{k_n b_n}{2}-\frac{1}{k_n}
\~,
\label{eqn:bQn}
\eeq
 where
\beq
        \bQxn{\lambda}
 \equiv \frac{1}{N_n} \sumiNn \abrac{\Qni - Q_n}^{\lambda}
\~.
\label{eqn:bQn_lambda}
\eeq
 Then the shifted point $Q_{s, n}$
 in the ansatz (\ref{eqn:dRdQn}),
 at which the leading systematic error
 due to the ansatz
 is minimal \cite{DMDDf1v},
 can be estimated by
\beq
   Q_{s, n}
 = Q_n + \frac{1}{k_n} \ln\bfrac{\sinh(k_n b_n/2)}{k_n b_n/2}
\~.
\label{eqn:Qsn}
\eeq
 Note that $Q_{s, n}$ differs
 from the central point of the $n$th bin, $Q_n$.
 Finally,
 substituting the ansatz (\ref{eqn:dRdQn})
 into Eq.~(\ref{eqn:f1v_dRdQ})
 and then letting $Q = Q_{s, n}$,
 we can obtain that
\beq
   f_{1, {\rm rec}}(v_{s, n})
 = \calN
   \bBigg{\frac{2 Q_{s, n} r_n}{F^2(Q_{s, n})}}
   \bbrac{\dd{Q} \ln \FQ \bigg|_{Q = Q_{s, n}} - k_n}
\~.
\label{eqn:f1v_Qsn}
\eeq
 Here
\(
   v_{s, n}
 = \alpha \sqrt{Q_{s, n}}
\),
 and the normalization constant $\calN$
 given in Eq.~(\ref{eqn:calN_int})
 can be estimated directly from the data by
\beq
   \calN
 = \frac{2}{\alpha}
   \bbrac{\sum_{a} \frac{1}{\sqrt{Q_a} \~ F^2(Q_a)}}^{-1}
\~,
\label{eqn:calN_sum}
\eeq
 where the sum runs over all events in the sample.

\subsection{Windowing the data set}
 In order to reduce the statistical uncertainty
 on the velocity distribution
 reconstructed by Eq.~(\ref{eqn:f1v_Qsn})
 and
 some uncontrolled systematic errors
 caused by neglecting terms of higher powers of $Q - Q_n$,
 as well as
 to offer a reasonable number of
 reconstructable velocity points $v_{s, n}$ of $f_1(v)$,
 it has been introduced in Ref.~\cite{DMDDf1v} that
 one can first collect experimental data
 in relatively small bins
 with linearly increased widths
 and then combining varying numbers of bins
 into overlapping ``windows''.
 Thus,
 we set that
 the bin widths satisfy
\(
   b_n
 = b_1 + (n - 1) \delta
\),
 Hence,
\beq
   Q_n
 = \Qmin + \abrac{n - \frac{1}{2}} b_1 + \bfrac{(n - 1)^2}{2} \delta
\~.
\label{eqn:Qn_delta}
\eeq
 Here the increment $\delta$ satisfies
\(
   \delta
 = 2 \aBig{\Qmax - \Qmin - B b_1} / B (B - 1)
\),
 $B$ being the total number of bins,
 and $Q_{\rm (min, max)}$ are
 the experimental minimal and maximal cut--off energies.
 Assume up to $n_W$ bins are collected into a window,
 with smaller windows at the borders of the range of $Q$.

 In order to distinguish the numbers of bins and windows,
 hereafter Latin indices $n,~m,~\cdots$ are used to label bins,
 and Greek indices $\mu,~\nu,~\cdots$ to label windows.
 For $1 \leq \mu \leq n_W$,
 the $\mu$th window simply consists of the first $\mu$ bins;
 for $n_W \leq \mu \leq B$,
 the $\mu$th window consists of bins
 $\mu-n_W + 1,~\mu-n_W + 2,~\cdots,~\mu$;
 and for $B \leq \mu \leq B+n_W-1$,
 the $\mu$th window consists of the last $n_W - (\mu - B)$ bins.
 This can also be described by introducing
 the indices $n_{\mu-}$ and $n_{\mu+}$
 which label the first and last bins
 contributing to the $\mu$th window,
 with
\cheqna
\beq
\renewcommand{\arraystretch}{1.3}
   n_{\mu-}
 = \cleft{\begin{array}{l c l}
           1,         & ~~~~~~ & {\rm for}~\mu \leq n_W, \\
           \mu-n_W+1, &        & {\rm for}~\mu \geq n_W,
          \end{array}}
\label{eqn:n_mu_minus}
\eeq
 and
\cheqnb
\beq
\renewcommand{\arraystretch}{1.3}
   n_{\mu+}
 = \cleft{\begin{array}{l c l}
           \mu, & ~~~~~~ & {\rm for}~\mu \leq B, \\
           B,   &        & {\rm for}~\mu \geq B.
          \end{array}}
\label{eqn:n_mu_plus}
\eeq
\cheqn
 The total number of windows
 defined through Eqs.~(\ref{eqn:n_mu_minus}) and (\ref{eqn:n_mu_plus})
 is evidently $W = B + n_W - 1$,
 i.e.~$1 \leq \mu \leq B + n_W - 1$.

 For a ``windowed'' data set,
 one can easily calculate
 the number of events per window as
\beq
   N_{\mu}
 = \sum_{n = n_{\mu-}}^{n_{\mu+}} N_n
\~,
\label{eqn:N_mu}
\eeq
 as well as,
 the average value of the measured recoil energies
\beq
   \Bar{Q - Q_{\mu}}|_{\mu}
 = \frac{1}{N_{\mu}}
   \abrac{\sum_{n = n_{\mu-}}^{n_{\mu+}} N_n \Bar{Q}|_{n}} - Q_{\mu}
\~,
\label{eqn:wQ_mu}
\eeq
 where $Q_{\mu}$ is the central point of the $\mu$th window.
 The exponential ansatz in Eq.~(\ref{eqn:dRdQn})
 is now assumed to hold over an entire window.
 We can then estimate the prefactor as
\(
   r_{\mu}
 = N_{\mu} / w_{\mu}
\),
 $w_{\mu}$ being the width of the $\mu$th window.
 The logarithmic slope of the recoil spectrum
 in the $\mu$th window, $k_{\mu}$,
 as well as
 the shifted point $Q_{s, \mu}$
 (from the central point of each ``window'', $Q_{\mu}$)
 can be calculated as in Eqs.~(\ref{eqn:bQn}) and (\ref{eqn:Qsn})
 with ``bin'' quantities replaced by ``window'' quantities.
 Finally,
 the covariance matrix of
 the estimates of $f_1(v)$
 at adjacent values of $v_{s, \mu} = \alpha \sqrt{Q_{s, \mu}}$
 is given by%
\footnote{
 Note that
 contributions involving
 the statistical error on the estimator for $\calN$
 in Eq.~(\ref{eqn:calN_sum})
 should in principle also,
 but do not be included here.
}
\beqn
 \conti  {\rm cov}\aBig{f_{1, {\rm rec}}(v_{s, \mu}), f_{1, {\rm rec}}(v_{s, \nu})}
         \non\\
 \=      \bfrac{f_{1, {\rm rec}}(v_{s, \mu}) f_{1, {\rm rec}}(v_{s, \nu})}
               {r_{\mu} r_{\nu}}
         {\rm cov}\abrac{r_{\mu}, r_{\nu}}
       + \abrac{2 \calN}^2
         \bfrac{Q_{s, \mu} Q_{s, \nu} r_{\mu} r_{\nu}}{F^2(Q_{s, \mu}) F^2(Q_{s, \nu})}
         {\rm cov}\abrac{k_{\mu}, k_{\nu}}
         \non\\
 \conti  ~~~~~~~~~~~~ 
       - \calN
         \cbrac{  \bfrac{f_{1, {\rm rec}}(v_{s, \mu})}{r_{\mu}}
                  \bfrac{2 Q_{s, \nu} r_{\nu} }{F^2(Q_{s, \nu})}
                  {\rm cov}\abrac{r_{\mu}, k_{\nu}}
                + \aBig{\mu \lgetsto \nu}}
\~.
\label{eqn:cov_f1v_Qs_mu}
\eeqn
\subsection{Numerical results}

 In this section,
 we present reconstruction results
 of the one--dimensional WIMP velocity distribution
 {\em before} taking
 the modification of
 the normalization constant of $f_1(v)$.

 First of all,
 since the lighter the WIMP mass,
 the more problematic
 the non--negligible experimental threshold energy,
 a light input WIMP mass of \mbox{$\mchi = 25$ GeV}
 has been considered in
 our simulations.
 As discussed and shown
 in Ref.~\cite{DMDDmchi},
 with ${\cal O}(500)$ recorded events
 (in one data set),
 the (light) input WIMP masses
 can be reconstructed pretty precisely
 with only \mbox{$\sim$ 10\%} (a few GeV) statistical uncertainties
 (more simulation results
  can also be found in Ref.~\cite{DMDDf1v-Bayesian}).
 Thus,
 in our simulations
 demonstrated here and in the next section,
 the reconstructed WIMP mass $\mchi$
 involved in the coefficient $\alpha$
 for estimating the reconstructed points $v_{s, \mu}$
 as well as
 the normalization constant $\calN$
 has been assumed
 to be known precisely with a negligible uncertainty.

 As in Ref.~\cite{DMDDf1v},
 $\rmXA{Ge}{76}$ has been chosen
 as our detector material for reconstructing $f_1(v)$.%
\footnote{
 Note that,
 while
 for a WIMP mass of \mbox{${\cal O}(100)$ GeV},
 the transformation constant $\alpha$
 defined in Eq.~(\ref{eqn:alpha})
 is larger with $\rmXA{Si}{28}$ as the target nucleus
 than with $\rmXA{Ge}{76}$,
 for a WIMP mass of \mbox{${\cal O}(25)$ GeV},
 the transformation constant $\alpha$
 with
 $\rmXA{Ge}{76}$
 is larger.
}
 As in Refs.~\cite{DMDDf1v, DMDDf1v-Bayesian},
 the WIMP--nucleus cross section
 appearing in the expression (\ref{eqn:dRdQ})
 for the recoil spectrum $dR / dQ$
 has been assumed
 to be only spin--independent (SI),
 \mbox{$\sigmapSI = 10^{-9}$ pb},
 and
 the commonly used analytic form
 for the elastic nuclear form factor:
\beq
   F_{\rm SI}^2(Q)
 = \bfrac{3 j_1(q R_1)}{q R_1}^2 e^{-(q s)^2}
\label{eqn:FQ_WS}
\eeq
 has been adopted.
 Here $Q$ is the recoil energy
 transferred from the incident WIMP to the target nucleus,
 $j_1(x)$ is a spherical Bessel function,
\(
   q
 = \sqrt{2 m_{\rm N} Q}
\)
 is the transferred 3-momentum,
 for the effective nuclear radius we use
\(
   R_1
 = \sqrt{R_A^2 - 5 s^2}
\)
 with
\(
        R_A
 \simeq 1.2 \~ A^{1/3}~{\rm fm}
\)
 and a nuclear skin thickness
\(
        s
 \simeq 1~{\rm fm}
\).

 By taking into account
 the orbital motion of the Solar system around our Galaxy
 as well as
 that of the Earth around the Sun,
 the
 shifted Maxwellian velocity distribution of halo WIMPs
 has been given by
 \cite{SUSYDM96, DMDDf1v}:
\beq
   f_{1, \sh}(v)
 = \frac{1}{\sqrt{\pi}} \afrac{v}{v_0 \ve}
   \bBig{  e^{-(v - \ve)^2 / v_0^2}
         - e^{-(v + \ve)^2 / v_0^2}  }
\~.
\label{eqn:f1v_sh}
\eeq
 Here
 \mbox{$v_0 \simeq 220$ km/s}
 is the Solar orbital speed around the Galactic center,
 and
 $\ve$ is the time-–dependent
 Earth's velocity in the Galactic frame
 \cite{Freese88,
       SUSYDM96}:
\beq
   \ve(t)
 = v_0 \bbrac{1.05 + 0.07 \cos\afrac{2 \pi (t - t_{\rm p})}{1~{\rm yr}}}
\~,
\label{eqn:ve}
\eeq
 with $t_{\rm p} \simeq$ June 2nd,
 the date
 on which the velocity of the Earth relative to the WIMP halo is maximal%
\footnote{
 As usual,
 in all our simulations
 the time dependence of the Earth's velocity in the Galactic frame,
 the second term of $\ve(t)$,
 will be ignored,
 i.e.~$\ve = 1.05 \~ v_0$
 is used.
}.
 Additionally,
 a common maximal cut--off
 on the one--dimensional WIMP velocity distribution
 has been set as \mbox{$\vmax = 700$ km/s}.

\begin{figure}[t!]
\begin{center}
\includegraphics[width=15cm]{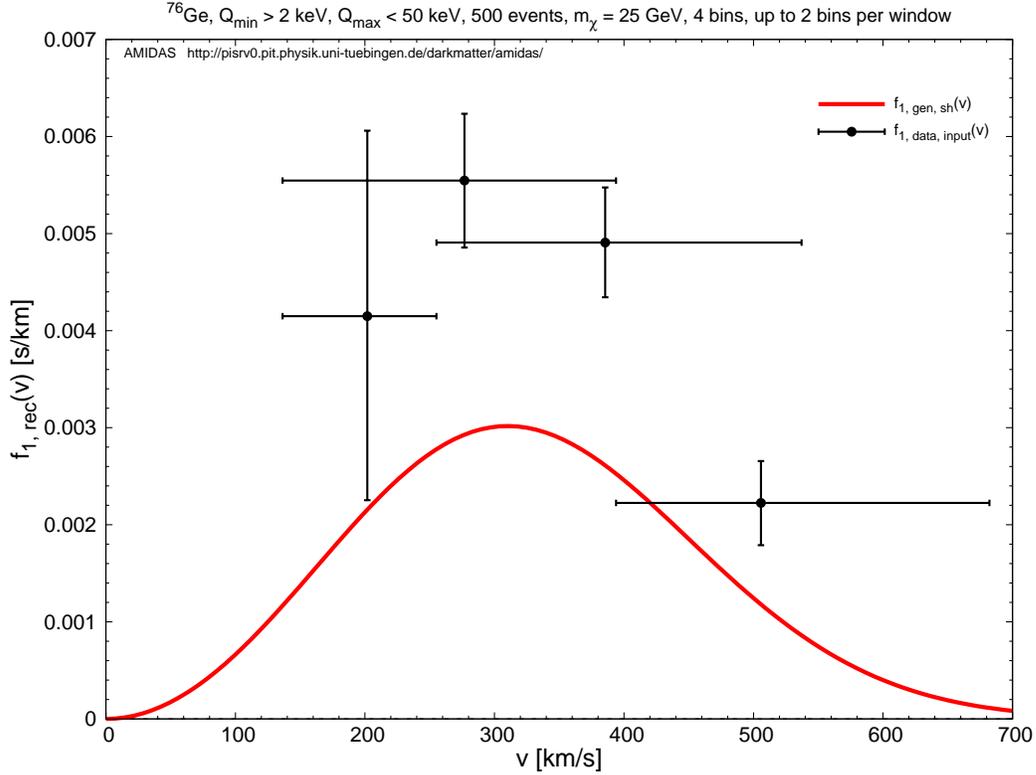} \\
\end{center}
\caption{
 The reconstructed
 (rough) velocity distribution (black crosses)
 with a $\rmXA{Ge}{76}$ target
 for an input WIMP mass of \mbox{$\mchi = 25$ GeV}
 and an experimental threshold energy
 of \mbox{$\Qmin = 2$ keV}.
 The vertical error bars show
 the square roots of the diagonal entries of the covariance matrix
 estimated by Eq.~(\ref{eqn:cov_f1v_Qs_mu}),
 whereas
 the horizontal bars indicate
 the sizes of the windows used
 for estimating $f_{1, {\rm rec}}(v_{s, \mu})$
 by Eq.~(\ref{eqn:f1v_Qsn}).
 The solid red curve
 is the generating shifted Maxwellian velocity distribution
 with an input value of \mbox{$v_0 = 220$ km/s}.
 See the text for further details.
}
\label{fig:f1v-I_0-Ge-025-0500-02-sh-sh_v0-flat-gen}
\end{figure}

 In our simulation shown
 in Fig.~\ref{fig:f1v-I_0-Ge-025-0500-02-sh-sh_v0-flat-gen},
 the experimental threshold energy
 has been set as
 \mbox{$\Qmin = 2$ keV}.
 Due to the maximal cut--off
 on the one--dimensional WIMP velocity distribution,
 $\vmax$,
 a kinematic maximal cut--off energy,
\beq
   Q_{\rm max, kin}
 = \frac{\vmax^2}{\alpha^2}
\~,
\label{eqn:Qmax_kin}
\eeq
 has to be considered.
 Since
 for our target $\rmXA{Ge}{76}$
 it is only \mbox{$Q_{\rm max, kin, Ge} = 52.65$ keV},
 the maximal experimental cut--off energy
 has been set to be only \mbox{$\Qmax = 50$ keV}.
 Meanwhile,
 since the lighter the WIMP mass,
 the steeper the expected recoil energy spectrum,
 \mbox{$b_1 = 5$ keV}
 width of the first energy bin
 has been used
 and
 the energy range between $\Qmin$ and $\Qmax$
 has been divided into
 only four bins ($B = 4$);
 up to two bins have been combined to a window
 and thus four windows ($W = 4$)%
\footnote{
 Note that
 the last window is neglected automatically
 in the \amidas\ code
 \cite{AMIDAS-web, AMIDAS-II},
 due to a very few expected event number
 in the last bin (window).
}
 will be reconstructed
 \cite{DMDDf1v-Bayesian}.
 Additionally,
 we assumed that
 all experimental systematic uncertainties
 as well as
 the uncertainty on the measurement of the recoil energy
 could be ignored.
 5,000 experiments with 500 total events on average%
\footnote{
 Note that,
 for our numerical simulations
 presented in this paper,
 the actual number of generated signals
 in each simulated experiment
 is Poisson--distributed around the expectation value.
}
 in one experiment have been simulated.

 In Fig.~\ref{fig:f1v-I_0-Ge-025-0500-02-sh-sh_v0-flat-gen},
 we show the reconstructed
 (rough) velocity distribution (black crosses)
 with a $\rmXA{Ge}{76}$ target.
 The vertical error bars show
 the square roots of the diagonal entries of the covariance matrix
 estimated by Eq.~(\ref{eqn:cov_f1v_Qs_mu}),
 whereas
 the horizontal bars indicate
 the sizes of the windows used
 for estimating $f_{1, {\rm rec}}(v_{s, \mu})$
 by Eq.~(\ref{eqn:f1v_Qsn}).
 As a comparison,
 the solid red curve
 indicates the generating shifted Maxwellian velocity distribution
 with an input value of \mbox{$v_0 = 220$ km/s}.
 It can be seen clearly that
 the reconstructed velocity distribution
 is around {\em two} times {\em overestimated}.
 This indicates in turn that,
 by using Eq.~(\ref{eqn:calN_sum}),
 the normalization constant $\calN$ of
 the velocity distribution function $f_1(v)$
 is {\em underestimated} as only the half!
 Remind that
 the experimental threshold energy used here
 is as low as only \mbox{2 keV}
 and the corresponding minimal cut--off of the velocity distribution
 is \mbox{134.44 km/s}
 (see also
  Fig.~\ref{fig:f1v-rmin_I_0-Ge-025-0500-02-sh-sh_v0-flat-gen}).

\section{Modification of the estimator for
         the normalization constant \boldmath$\calN$}

 In this section,
 we take into account the effect of
 a non--negligible experimental threshold energy
 ($\Qmin > 0$)
 and
 introduce a {\em model--independent} modification of
 the estimator for the normalization constant $\calN$
 step by step.

\subsection{Non--zero minimal cut--off velocity}

 First,
 we consider
 the minimal cut--off of the velocity distribution
 due to the non--zero experimental threshold energy,
 $\vmin(\Qmin) \equiv \vmin^{\ast}$,
 in the use of the normalization condition
 (\ref{eqn:normalization_infty}).
 From Eq.~(\ref{eqn:f1v_dRdQ}),
 since
\(
   v
 = \alpha \sqrt{Q}
\),
 by using integration by parts,
 we can obtain that
\beqn
     \int_{\vmin^{\ast}}^{\vmax} f_1(v) \~ dv
 \=  \calN
     \int_{\Qmin}^{\Qmax^{\ast}}
     \cbrac{-2 Q \cdot \ddRdQoFQdQ} \afrac{\alpha}{2 \sqrt{Q}} dQ
     \non\\
 \=  \calN \afrac{\alpha}{2}
     \cbrac{  \frac{2 \Qmin^{1 / 2}}{\FQmin} \adRdQ_{Q = \Qmin}
            + \int_{\Qmin}^{\Qmax^{\ast}}
              \frac{1}{\sqrt{Q}} \bdRdQoFQ dQ}
     \non\\
 \=  \calN \afrac{\alpha}{2}
     \bbrac{  \frac{2 \Qmin^{1 / 2} r(\Qmin)}{\FQmin}
            + I_0(\Qmin, \Qmax^{\ast})}
\~.
\label{eqn:int_vmin_vmax}
\eeqn
 Here we define
\(
        \Qmax^{\ast}
 \equiv {\rm min}\abrac{\Qmax,~Q_{\rm max, kin}}
\),
 the smaller one between
 the experimental and kinematic cut--off energies
 and can be understood as
 the upper bound of the recoil energy
 of the recorded events;
 $Q_{\rm (min, max)}$ are the experimental
 minimal and maximal cut--off energies.
 Since the WIMP--nucleus scattering spectrum
 is expected to be exponential,
 the term of $\abrac{dR / dQ}_{Q = \Qmax^{\ast}}$
 has been ignored here,
 whereas
\beq
        r(\Qmin)
 \equiv \adRdQ_{{\rm expt},\~Q = \Qmin}
 =      r_1 \~ e^{k_1 (\Qmin - Q_{s, 1})}
\label{eqn:rmin}
\eeq
 with
\(
   r_1
 = N_1 / b_1
\),
 an estimated value
 of the {\em measured} recoil spectrum
 $(dR / dQ)_{\rm expt}$ ({\em before}
 the normalization by the exposure $\cal E$) at $Q = \Qmin$.
 Finally,
 $I_n(\Qmin, \Qmax^{\ast})$ can be estimated through the sum
 running over all events in the data set that satisfy
 $Q_a \in [\Qmin, \Qmax^{\ast}]$:
\beq
        I_n(\Qmin, \Qmax^{\ast})
 \equiv \int_{\Qmin}^{\Qmax^{\ast}} Q^{(n - 1) / 2} \bdRdQoFQ dQ
 \to    \sum_a \frac{Q_a^{(n - 1) / 2}}{F^2(Q_a)}
\~.
\label{eqn:In_sum}
\eeq

 Once we neglect the (small) contributions
 from both $v > \vmax$ ($< 0.5\%$)
 and $v \le \vmin^{\ast}$
 and approximate the normalization condition by
\beq
         \int_{\vmin^{\ast}}^{\vmax} f_1(v) \~ dv
 \approx  1
\~,
\label{eqn:normalization_vmin_vmax}
\eeq
 a modified estimator for
 the normalization constant can be given
 from Eq.~(\ref{eqn:int_vmin_vmax}) as
\beq
     \calN
 \approx
     \frac{2}{\alpha}
     \bbrac{  \frac{2 \Qmin^{1 / 2} r(\Qmin)}{\FQmin}
            + I_0(\Qmin, \Qmax^{\ast})}^{-1}
\~.
\label{eqn:calN_sum_rmin_I0}
\eeq
 Remind that
 the first (extra) term in the bracket is caused by
 the non--zero minimal cut--off velocity \mbox{$\vmin^{\ast}$}
 and vanishes
 once the threshold energy is negligible (\mbox{$\Qmin \simeq 0$}).

\begin{figure}[t!]
\begin{center}
\includegraphics[width=15cm]{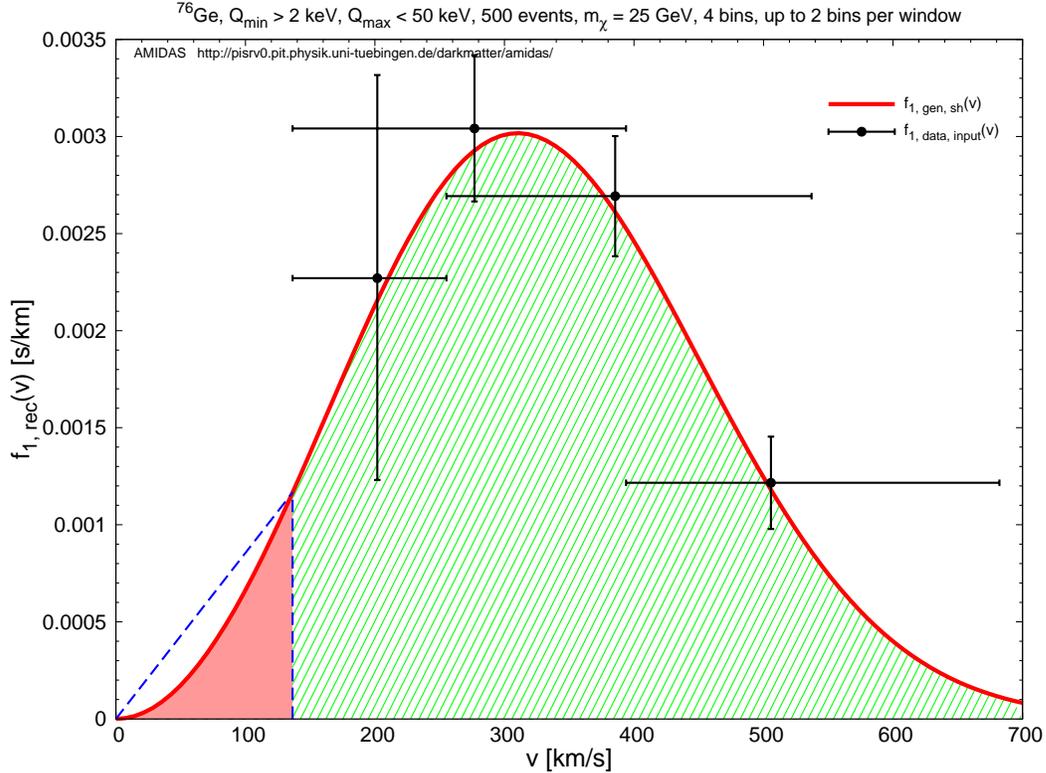} \\
\end{center}
\caption{
 The reconstructed
 (rough) velocity distribution (black crosses)
 with the normalization constant
 estimated by Eq.~(\ref{eqn:calN_sum_rmin_I0}).
 The vertical dashed blue line indicates
 \mbox{$\vmin(\Qmin = 2~{\rm keV}) = 136.44$ km/s}.
 All parameters are as
 in Fig.~\ref{fig:f1v-I_0-Ge-025-0500-02-sh-sh_v0-flat-gen}.
 See the text for detailed discussions.
}
\label{fig:f1v-rmin_I_0-Ge-025-0500-02-sh-sh_v0-flat-gen}
\end{figure}

 In Fig.~\ref{fig:f1v-rmin_I_0-Ge-025-0500-02-sh-sh_v0-flat-gen},
 we show
 the reconstructed
 (rough) velocity distribution (black crosses)
 with the normalization constant
 estimated by Eq.~(\ref{eqn:calN_sum_rmin_I0}).
 The vertical dashed blue line indicates
 \mbox{$\vmin(\Qmin = 2~{\rm keV}) = 136.44$ km/s}.
 Thus
 the meshed green area above
 \mbox{$\vmin^{\ast} = 136.44$ km/s}
 denotes the integral
 on the left--hand side of the normalization condition
 (\ref{eqn:normalization_vmin_vmax}),
 whereas
 the shaded light--red area blow $\vmin^{\ast}$
 has been neglected.

 Fig.~\ref{fig:f1v-rmin_I_0-Ge-025-0500-02-sh-sh_v0-flat-gen}
 shows that
 the reconstructed
 velocity distribution
 with the modified normalization constant
 given by Eq.~(\ref{eqn:calN_sum_rmin_I0})
 is strongly improved and
 could already match the true (input) velocity distribution
 (the solid red curve)
 pretty well.
 However,
 Fig.~\ref{fig:f1v-rmin_I_0-Ge-025-0500-02-sh-sh_v0-flat-gen}
 shows also clearly that
 the contribution
 below the non--zero minimal cut--off velocity $\vmin^{\ast}$
 (the shaded light--red area)
 would still be non--negligible!

\subsection{Contribution below the non--zero minimal cut--off velocity}

 The reconstructed velocity distribution
 shown in Fig.~\ref{fig:f1v-rmin_I_0-Ge-025-0500-02-sh-sh_v0-flat-gen}
 matches almost perfectly to the true (input) distribution.
 However,
 as pointed out in the previous section,
 since
 not only the tiny contribution from the $v > \vmax$ area
 ($< 0.5\%$)
 but also a pretty large part
 from \mbox{$v \le \vmin^{\ast} = 136.44$ km/s}
 (the shaded light--red area)
 has been omitted,
 the normalization condition (\ref{eqn:normalization_vmin_vmax})
 with the integral over $f_1(v)$
 only between $\vmin^{\ast}$ and $\vmax$
 could be considerably underestimated.
 Hence,
 an estimator for
 the area under the (true) velocity distribution function
 in the velocity range $[0, \vmin^{\ast}]$
 should be given.

 In this section,
 we suggest a {\em model--independent} estimate
 for the area below \mbox{$\vmin^{\ast}$}.
 As sketched in Fig.~\ref{fig:f1v-rmin_I_0-Ge-025-0500-02-sh-sh_v0-flat-gen},
 for this aim,
 the value of the velocity distribution function at $v = \vmin^{\ast}$
 has to be estimated
 and we approximate then the area of $v \le \vmin^{\ast}$ simply
 by a triangle.
 We start with
 the reconstructed recoil spectrum in the first $Q-$bin
 which can be given by
\beq
         \adRdQ_{{\rm expt}, \~ 1}
 =       r_1 \~ e^{k_1 (Q - Q_{s, 1})}
\~.
\label{eqn:dRdQ1}
\eeq
 By using Eq.~(\ref{eqn:f1v_dRdQ}),
 an expression,
 similar to Eq.~(\ref{eqn:f1v_Qsn}),
 for the value of
 the reconstructed velocity distribution function
 at $v = \vmin^{\ast}$
 can be found as
\beq
    f_{1, {\rm rec}}(\vmin^{\ast})
  = \calN
    \bBigg{\frac{2 \Qmin r(\Qmin)}{\FQmin}}
    \bbrac{\dd{Q} \ln \FQ \bigg|_{Q = \Qmin} - k_1}
  \equiv
    \calN \td{f}_{1, {\rm rec}}(\vmin^{\ast})
\~.
\label{eqn:f1v_Qmin}
\eeq
 Combining Eqs.~(\ref{eqn:int_vmin_vmax}) and (\ref{eqn:f1v_Qmin}),
 the integral over $f_1(v)$
 between 0 and $\vmax^{\ast}$ can be given by
\beqn
      \intz f_1(v) \~ dv
 \eqnsimeq
      \abrac{\int_0^{\vmin^{\ast}} + \int_{\vmin^{\ast}}^{\vmax}} f_1(v) \~ dv
      \non\\
 \eqnsimeq
      \calN \tilde{f}_{1, {\rm rec}}(\vmin^{\ast}) \cdot \frac{\vmin^{\ast}}{2}
    + \calN \afrac{\alpha}{2}
      \bbrac{  \frac{2 \Qmin^{1 / 2} r(\Qmin)}{\FQmin}
              + I_0(\Qmin, \Qmax^{\ast})}
      \non\\
 \=   \calN \afrac{\alpha}{2}
      \bbrac{  \tilde{f}_{1, {\rm rec}}(\vmin^{\ast}) \sqrt{\Qmin}
             + \frac{2 \Qmin^{1 / 2} r(\Qmin)}{\FQmin}
             + I_0(\Qmin, \Qmax^{\ast})}
      \non\\
 \=   1
\~.
\label{eqn:normalization_infty_mod}
\eeqn
 Therefore,
 we can obtain
 the model--independent approximation for
 the normalization constant of
 the reconstructed WIMP velocity distribution as
\beq
    \calN
 =  \frac{2}{\alpha}
    \bbrac{  \tilde{f}_{1, {\rm rec}}(\vmin^{\ast}) \~ \Qmin^{1 / 2}
           + \frac{2 \Qmin^{1 / 2} r(\Qmin)}{\FQmin}
           + I_0(\Qmin, \Qmax^{\ast})}^{-1}
\~.
\label{eqn:calN_sum_mod}
\eeq
 Note that
 the modified normalization constant $\calN$
 given here
 depends now on the estimates of $r_1$ and $k_1$.%
\footnote{
 In fact,
 the normalization constant $\calN$
 given in Eq.~(\ref{eqn:calN_sum_rmin_I0})
 depends also on the estimates of $r_1$ and $k_1$.
 However,
 note that,
 we {\em didn't} consider
 a modification of the covariance matrix
 used for estimating the statistical uncertainty bars
 shown in Fig.~\ref{fig:f1v-rmin_I_0-Ge-025-0500-02-sh-sh_v0-flat-gen}
 as well as in the upper frames of
 Figs.~\ref{fig:f1v-Ge-025-0500-05-sh-sh_v0-flat-gen} and
 Figs.~\ref{fig:f1v-Si-025-0500-05-sh-sh_v0-flat-gen},
 since Eq.~(\ref{eqn:calN_sum_rmin_I0})
 is only an intermediate product
 for our final expression (\ref{eqn:calN_sum_mod}).
}
 Hence,
 the covariance matrix of the estimates $f_{1, {\rm rec}}(v_{s, \mu})$
 needs to be modified as%
\footnote{
 Here we neglect again
 the relatively much smaller correlation between
 the uncertainties on $I_0$ and $r_1$, $k_1$.
}:
\beqn
 \conti  {\rm cov}\aBig{f_{1, {\rm rec}}(v_{s, \mu}), f_{1, {\rm rec}}(v_{s, \nu})}
         \non\\
 \=      \bfrac{f_{1, {\rm rec}}(v_{s, \mu}) f_{1, {\rm rec}}(v_{s, \nu})}
               {r_{\mu} r_{\nu}}
         {\rm cov}\abrac{r_{\mu}, r_{\nu}}
       + \abrac{2 \calN}^2
         \bfrac{Q_{s, \mu} Q_{s, \nu} r_{\mu} r_{\nu}}{F^2(Q_{s, \mu}) F^2(Q_{s, \nu})}
         {\rm cov}\abrac{k_{\mu}, k_{\nu}}
         \non\\
 \conti  ~~~~
       - \calN
         \cbrac{  \bfrac{f_{1, {\rm rec}}(v_{s, \mu})}{r_{\mu}}
                  \bfrac{2 Q_{s, \nu} r_{\nu} }{F^2(Q_{s, \nu})}
                  {\rm cov}\abrac{r_{\mu}, k_{\nu}}
                + \aBig{\mu \lgetsto \nu}}
         \non\\
 \conti  ~~~~ ~~~~ 
       + \bPp{f_{1, {\rm rec}}(v_{s, \mu})}{r_1}
         \bPp{f_{1, {\rm rec}}(v_{s, \nu})}{r_1}
         \sigma^2(r_1)
       + \bPp{f_{1, {\rm rec}}(v_{s, \mu})}{k_1}
         \bPp{f_{1, {\rm rec}}(v_{s, \nu})}{k_1}
         \sigma^2(k_1)
         \non\\
 \conti  ~~~~ ~~~~ ~~~~ 
       + \cbrac{  \bPp{f_{1, {\rm rec}}(v_{s, \mu})}{r_1}
                  \bPp{f_{1, {\rm rec}}(v_{s, \nu})}{k_1}
                + \aBig{\mu \lgetsto \nu}  }
         {\rm cov}\abrac{r_1, k_1}
         \non\\
 \conti  ~~~~ ~~~~ ~~~~ ~~~~ 
       + \cleft{   \bPp{f_{1, {\rm rec}}(v_{s, \mu})}{r_1}
                   \bPp{f_{1, {\rm rec}}(v_{s, \nu})}{r_{\nu}}
                   {\rm cov}\abrac{r_1, r_{\nu}}  }
         \non\\
 \conti  ~~~~ ~~~~ ~~~~ ~~~~ ~~~~ ~~ 
                 + \bPp{f_{1, {\rm rec}}(v_{s, \mu})}{r_1}
                   \bPp{f_{1, {\rm rec}}(v_{s, \nu})}{k_{\nu}}
                   {\rm cov}\abrac{r_1, k_{\nu}}
         \non\\
 \conti  ~~~~ ~~~~ ~~~~ ~~~~ ~~~~ ~~~~ 
       +           \bPp{f_{1, {\rm rec}}(v_{s, \mu})}{k_1}
                   \bPp{f_{1, {\rm rec}}(v_{s, \nu})}{r_{\nu}}
                   {\rm cov}\abrac{k_1, r_{\nu}}
         \non\\
 \conti  ~~~~ ~~~~ ~~~~ ~~~~ ~~~~ ~~~~ ~~ 
       + \cBiggr{  \bPp{f_{1, {\rm rec}}(v_{s, \mu})}{k_1}
                   \bPp{f_{1, {\rm rec}}(v_{s, \nu})}{k_{\nu}}
                   {\rm cov}\abrac{k_1, k_{\nu}}
                 + \aBig{\mu \lgetsto \nu}  }
\~.
\label{eqn:cov_f1v_Qs_mu_mod}
\eeqn
 All derivatives of $f_{1, {\rm rec}}(v_{s, \mu})$
 to $r_1$, $r_{\mu}$, $k_1$ and $k_{\mu}$ needed here
 are given in Appendix A.2.

\begin{figure}[t!]
\begin{center}
\includegraphics[width=15cm]{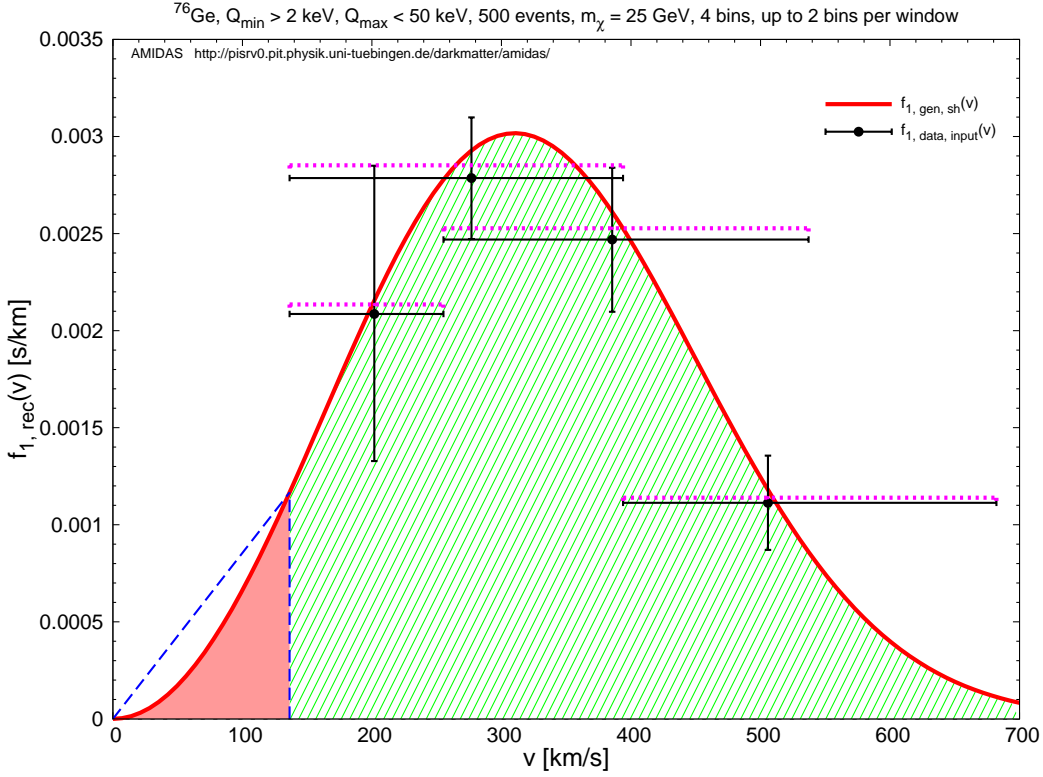} \\
\end{center}
\caption{
 The reconstructed (rough) velocity distribution
 (black crosses)
 with the normalization constant
 estimated by Eq.~(\ref{eqn:calN_sum_mod})
 and the modified statistical uncertainties
 given by Eq.~(\ref{eqn:cov_f1v_Qs_mu_mod}).
 The dotted magenta horizontal lines
 indicate the reconstructed velocity distributions
 including the correction of the theoretical estimate of
 the difference between the triangular approximation
 and the integral
 between 0 and $\vmin^{\ast}$
 (see the next section).
 All parameters are as
 in Fig.~\ref{fig:f1v-I_0-Ge-025-0500-02-sh-sh_v0-flat-gen}.
}
\label{fig:f1v-Ge-025-0500-02-sh-sh_v0-flat-gen}
\end{figure}

 In Fig.~\ref{fig:f1v-Ge-025-0500-02-sh-sh_v0-flat-gen},
 we show
 the reconstructed (rough) velocity distribution
 (black crosses)
 with the normalization constant
 estimated by Eq.~(\ref{eqn:calN_sum_mod})
 and the modified statistical uncertainties
 given by Eq.~(\ref{eqn:cov_f1v_Qs_mu_mod}).
 It can be found that,
 due to the contribution of the first term
 in the bracket in Eq.~(\ref{eqn:calN_sum_mod}),
 the reconstructed velocity distribution points
 are now a little bit underestimated.
 Meanwhile,
 the statistical uncertainties
 (vertical bars)
 given by Eq.~(\ref{eqn:cov_f1v_Qs_mu_mod})
 becomes now a bit larger,
 except the first one,
 which is significantly reduced.

\begin{figure}[p!]
\begin{center}
\includegraphics[width=15cm]{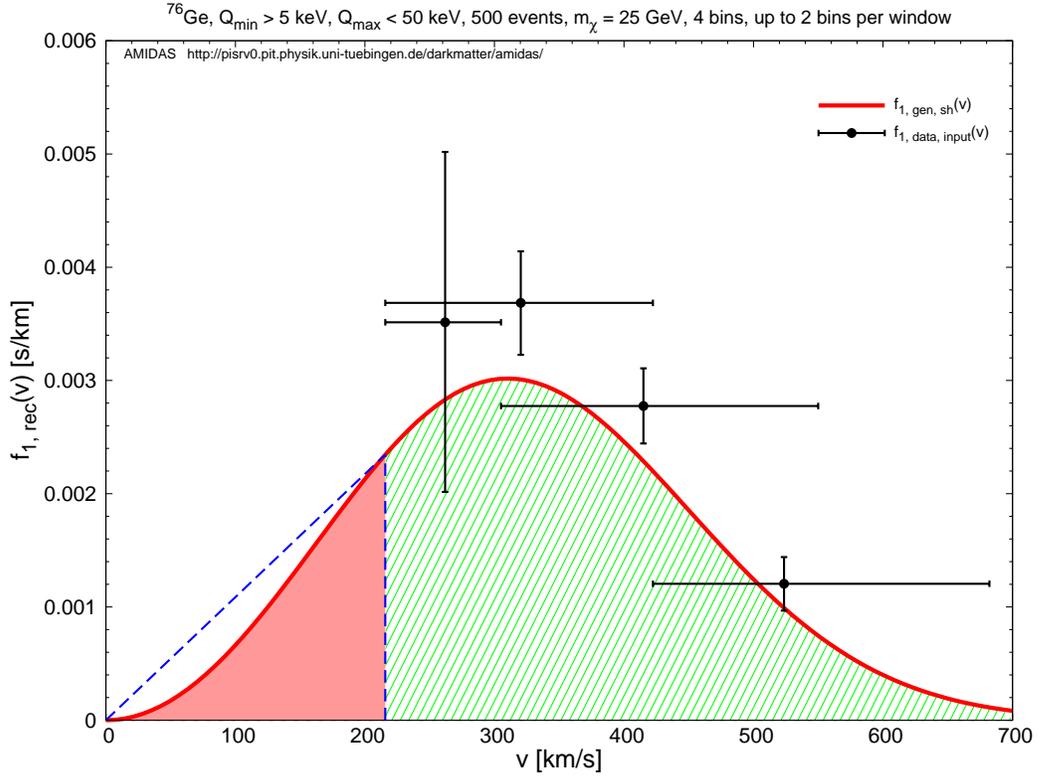}  \\ \vspace{1.5cm}
\includegraphics[width=15cm]{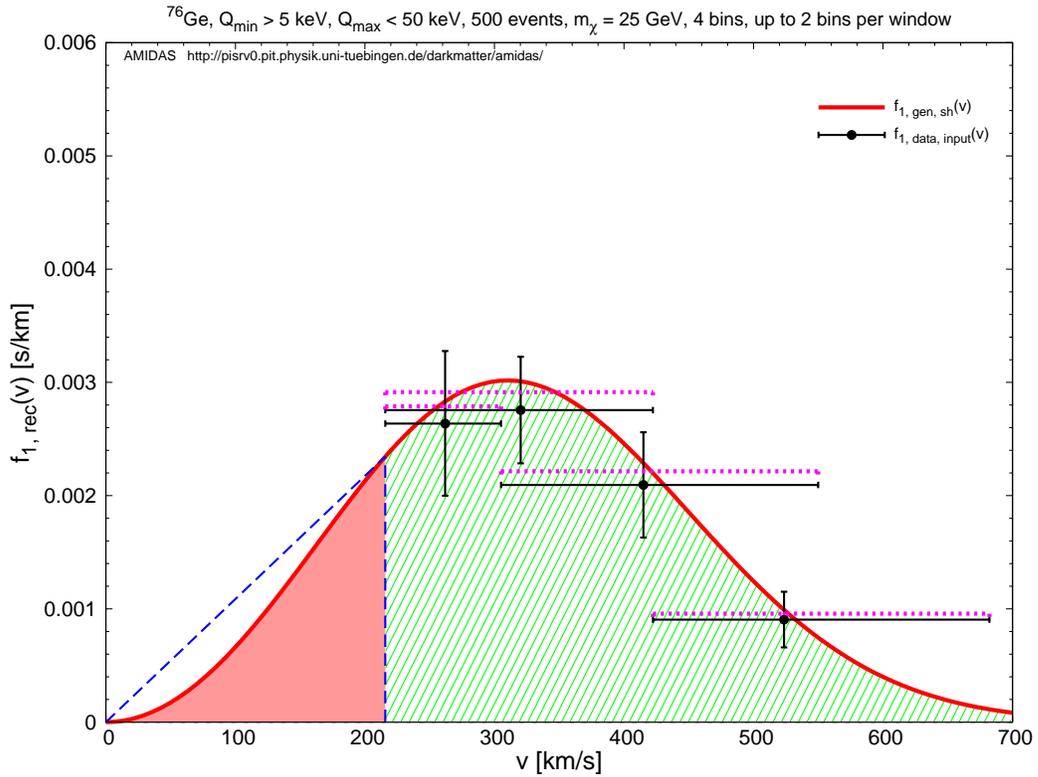}           \\
\vspace{-0.25cm}
\end{center}
\caption{
 As in Figs.~\ref{fig:f1v-rmin_I_0-Ge-025-0500-02-sh-sh_v0-flat-gen} (upper)
 and \ref{fig:f1v-Ge-025-0500-02-sh-sh_v0-flat-gen} (lower),
 except that
 the experimental threshold energy
 has been increased to $\Qmin = 5$ keV.
 Note that
 the vertical scale of $f_{1, {\rm rec}}(v)$ is different here.
}
\label{fig:f1v-Ge-025-0500-05-sh-sh_v0-flat-gen}
\end{figure}
\begin{figure}[p!]
\begin{center}
\includegraphics[width=15cm]{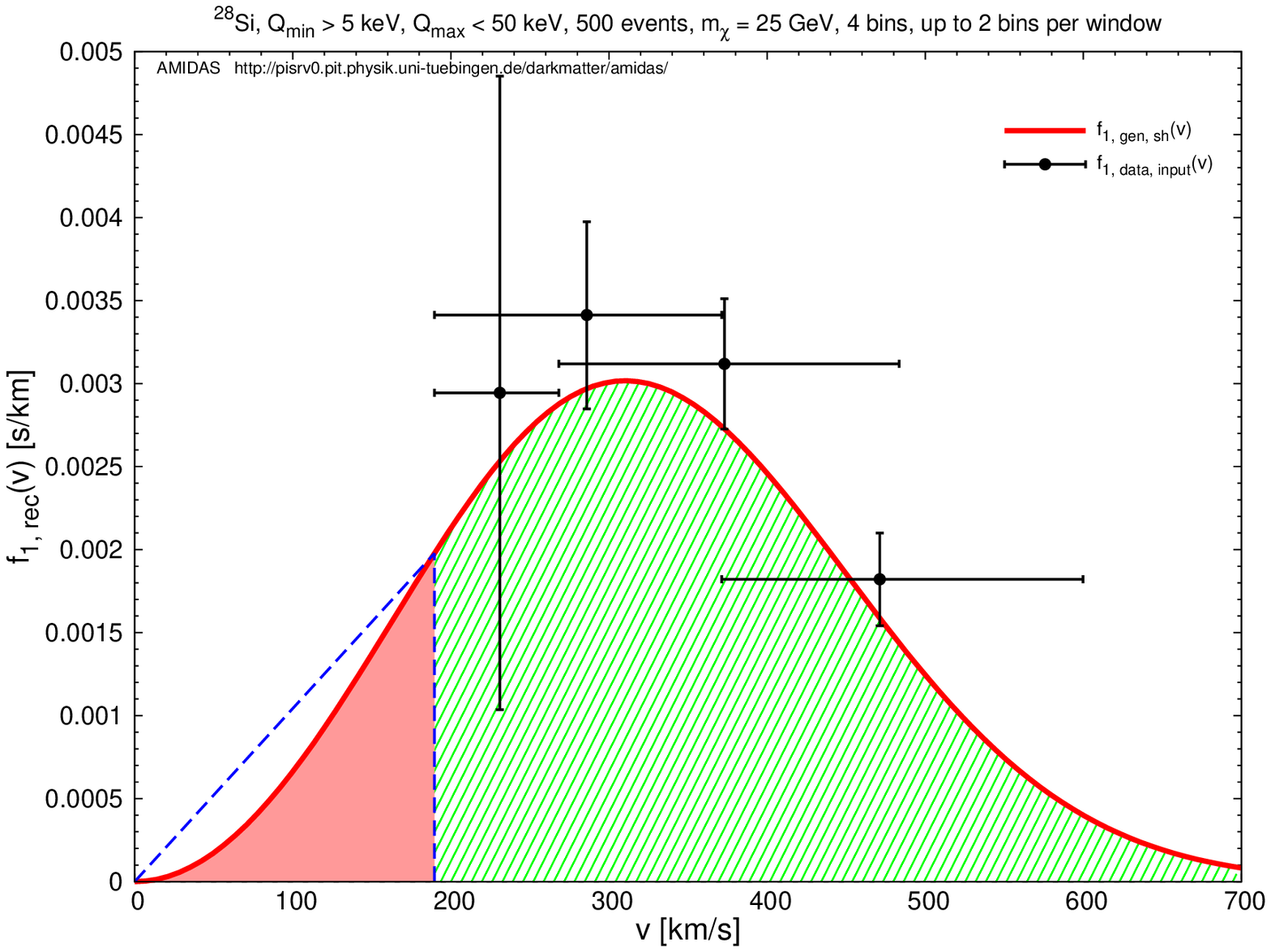}  \\ \vspace{1.5cm}
\includegraphics[width=15cm]{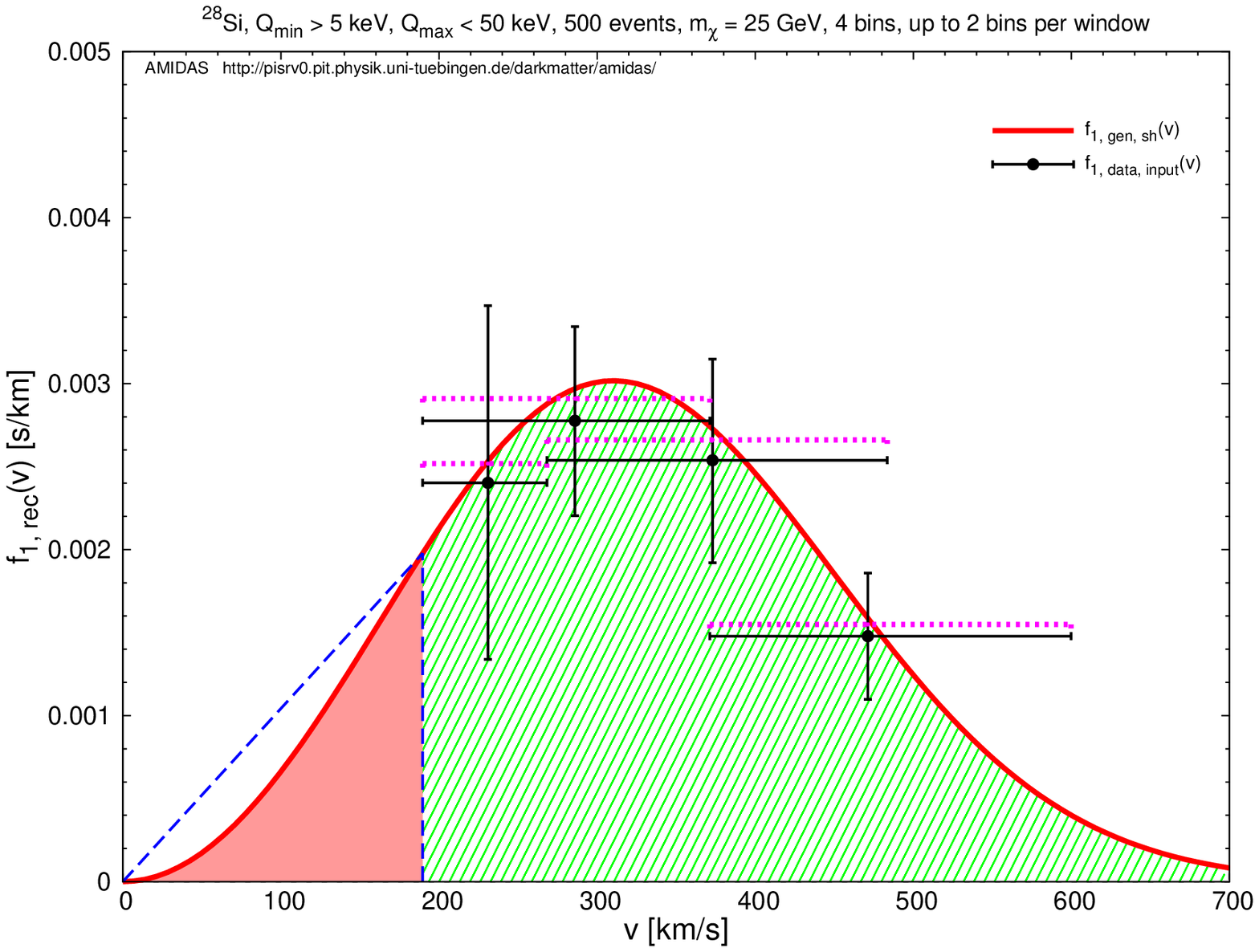}           \\
\vspace{-0.25cm}
\end{center}
\caption{
 As in Figs.~\ref{fig:f1v-Ge-025-0500-05-sh-sh_v0-flat-gen},
 except that
 $\rmXA{Si}{28}$ is used as the target nucleus.
 Note that
 the vertical scale of $f_{1, {\rm rec}}(v)$ is different here.
}
\label{fig:f1v-Si-025-0500-05-sh-sh_v0-flat-gen}
\end{figure}

 As a stricter check of our model--independent modification of
 the normalization constant $\calN$,
 in Figs.~\ref{fig:f1v-Ge-025-0500-05-sh-sh_v0-flat-gen},
 we increase the experimental threshold energy to
 $\Qmin = 5$ keV.
 For a germanium detector,
 the corresponding minimal cut--off velocity of the incident WIMPs
 is now \mbox{215.72
 km/s}
 and thus
 the contribution from the area of $v \le \vmin^{\ast}$
 becomes much larger.
 Hence,
 the underestimate of the normalization constant
 given by Eq.~(\ref{eqn:calN_sum_rmin_I0})
 and in turn the overestimate of
 the reconstructed velocity distribution points
 are more clear.
 In contrast,
 by using Eq.~(\ref{eqn:calN_sum_mod})
 the reconstructed points
 can still match the true (input) velocity distribution
 pretty well.
 However,
 and more importantly,
 the systematic bias
 between the triangular approximation
 and the real value of the integral of the $v \le \vmin^{\ast}$ area
 becomes more obviously and problematic.
 We will discuss this drawback
 in more details
 in the next section.

 Finally,
 in Figs.~\ref{fig:f1v-Si-025-0500-05-sh-sh_v0-flat-gen},
 a lighter target nucleus
 $\rmXA{Si}{28}$ is used as the detector material.
 For a silicon detector
 with an experimental threshold energy of 5 keV,
 the corresponding minimal cut--off velocity of the incident WIMPs
 is \mbox{189.64 km/s}.
 As the case with a $\rmXA{Ge}{76}$ target,
 the estimator (\ref{eqn:calN_sum_mod})
 for the normalization constant $\calN$
 combined with the modified covariance matrix
 given in Eq.~(\ref{eqn:cov_f1v_Qs_mu_mod})
 can not only correct
 the overestimated reconstructed velocity distribution
 very well,
 but also
 strongly reduce the large statistical uncertainty
 on the first reconstructed distribution point
 $f_{1, {\rm rec}}(v_{s, 1})$,
 although
 the statistical uncertainties
 are (a bit) larger than those
 given with the Ge target
 (cf.~Figs.~\ref{fig:f1v-Ge-025-0500-05-sh-sh_v0-flat-gen}).

\subsection{Bias of the estimator (\ref{eqn:calN_sum_mod}) for
            the normalization constant \boldmath$\calN$}

 As revealed
 in Figs.~\ref{fig:f1v-rmin_I_0-Ge-025-0500-02-sh-sh_v0-flat-gen}
 to \ref{fig:f1v-Si-025-0500-05-sh-sh_v0-flat-gen},
 our model--independent triangular estimator
 for the integral over $f_1(v)$ between 0 and $\vmin^{\ast}$
 is somehow overestimated.
 Then
 the normalization constant $\calN$ and
 the reconstructed velocity distribution points
 are in turn (a bit) underestimated.
 In this section,
 we discuss therefore
 the bias of
 the estimator (\ref{eqn:calN_sum_mod}) for
 the normalization constant $\calN$
 in more details.

\begin{table}[p!]
\begin{center}
\renewcommand{\arraystretch}{2.2}
\setlength{\tabcolsep}{3pt}
{\footnotesize
\begin{tabular}{|| c | c | c | c | c | c | c ||}
\hline
\hline
 \multirow{2}{*}{\makebox[2.1 cm][c]{$\vmin^{\ast}$ [km/s]}}  &
 \multicolumn{4}{| c |}{$\Qmin$ [keV]} &
 \multirow{2}{*}{\makebox[3.1 cm][c]
 {\large $\frac{\Delta_0^{\vmin^{\ast}}}{\int_0^{\vmin^{\ast}} f_1(v) \~ dv}$}} &
 \multirow{2}{*}{\makebox[4.8 cm][c]
 {{\large $\frac{\Delta_0^{\vmin^{\ast}} - \int_0^{\vmin^{\ast}} f_1(v) \~ dv}
                {\int_0^{\vmax} f_1(v) \~ dv}$}~~[\%]}}                         \\
\cline{2-5}
                                        &
 \makebox[1.35 cm][c]{$\rmXA{Si}{ 28}$} &
 \makebox[1.35 cm][c]{$\rmXA{Ar}{ 40}$} &
 \makebox[1.35 cm][c]{$\rmXA{Ge}{ 76}$} &
 \makebox[1.35 cm][c]{$\rmXA{Xe}{136}$} &
                                        & \\
\hline
  10 & ~0.014 & ~0.013 & 0.011 & 0.008 & 1.4997 & 0.0012 \\
\hline
  20 & ~0.056 & ~0.054 & 0.043 & 0.031 & 1.4987 & 0.0094 \\
\hline
  30 & ~0.125 & ~0.120 & 0.097 & 0.069 & 1.4970 & 0.0315 \\
\hline
  40 & ~0.222 & ~0.214 & 0.172 & 0.123 & 1.4946 & 0.0742 \\
\hline
  50 & ~0.348 & ~0.334 & 0.269 & 0.192 & 1.4916 & 0.1435 \\
\hline
  60 & ~0.500 & ~0.482 & 0.387 & 0.276 & 1.4877 & 0.2452 \\
\hline
  70 & ~0.681 & ~0.655 & 0.526 & 0.376 & 1.4830 & 0.3839 \\
\hline
  80 & ~0.890 & ~0.856 & 0.688 & 0.491 & 1.4775 & 0.5635 \\
\hline
  90 & ~1.126 & ~1.083 & 0.870 & 0.621 & 1.4711 & 0.7868 \\
\hline
 100 & ~1.390 & ~1.338 & 1.074 & 0.767 & 1.4637 & 1.0551 \\
\hline
 120 & ~2.002 & ~1.926 & 1.547 & 1.105 & 1.4458 & 1.7246 \\
\hline
 140 & ~2.725 & ~2.622 & 2.106 & 1.504 & 1.4234 & 2.5495 \\
\hline
 160 & ~3.560 & ~3.424 & 2.750 & 1.964 & 1.3960 & 3.4757 \\
\hline
 180 & ~4.504 & ~4.334 & 3.481 & 2.486 & 1.3634 & 4.4152 \\
\hline
 200 & ~5.561 & ~5.350 & 4.298 & 3.069 & 1.3253 & 5.2473 \\
\hline
 220 & ~6.729 & ~6.474 & 5.200 & 3.713 & 1.2816 & 5.8242 \\
\hline
 240 & ~8.008 & ~7.705 & 6.189 & 4.419 & 1.2323 & 5.9807 \\
\hline
 260 & ~9.398 & ~9.042 & 7.263 & 5.186 & 1.1777 & 5.5473 \\
\hline
 280 & 10.900 & 10.487 & 8.423 & 6.015 & 1.1180 & 4.3668 \\
\hline
 300 & 12.512 & 12.039 & 9.670 & 6.905 & 1.0537 & 2.3101 \\
\hline
\hline
\end{tabular}
}
\caption{
 The theoretically estimated ratios of the triangular approximation
 (to the integral over $f_1(v)$ in the range
  between 0 and $\vmin^{\ast}$)
 to the integral itself
 as well as
 the fractions of the difference between
 the triangular approximation
 and the integral
 between 0 and $\vmin^{\ast}$
 (i.e.~the little overestimated amount)
 to the integral in the entire velocity range
 between 0 and $\vmax$.
 The most commonly used model
 for the Galactic WIMP velocity distribution function,
 $f_{1, \sh}(v)$ given in Eq.~(\ref{eqn:f1v_sh}),
 has been used.
 20 different values of $\vmin^{\ast}$
 from 10 to 300 km/s
 for four most commonly used detector materials
 are given here.
 The $\Qmin$ values have been
 estimated for the WIMP mass of \mbox{$\mchi = 25$ GeV}.
}
\label{tab:bias_calN_sum_mod}
\end{center}
\end{table}

 Taking the most commonly used
 shifted Maxwellian velocity distribution
 $f_{1, \sh}(v)$ given in Eq.~(\ref{eqn:f1v_sh})
 as our theoretical assumption%
\footnote{
 Recently,
 several modifications of the Maxwellian velocity distribution
 have been introduced
 (see e.g.~Refs.~\cite{Lisanti10, YYMao, Kuhlen13}).
 However,
 as described in the papers,
 the significant differences between these (empirical) expressions
 and the shifted Maxwellian velocity distribution $f_{1, \sh}(v)$
 are only in the high--velocity tail.
},
 in Table \ref{tab:bias_calN_sum_mod}
 we give
 the
 estimated ratios of the triangular approximation
 (to the integral over $f_1(v)$ in the range
  between 0 and $\vmin^{\ast}$)
 to the integral itself:
\beq
   \frac{\Delta_0^{\vmin^{\ast}}}{\int_0^{\vmin^{\ast}} f_1(v) \~ dv}
\~,
\eeq
 as well as
 the fractions of the difference between
 the triangular approximation
 and the integral
 between 0 and $\vmin^{\ast}$
 (i.e.~the little overestimated amount)
 to the integral in the entire velocity range
 between 0 and $\vmax$:
\beq
   \frac{\Delta_0^{\vmin^{\ast}} - \int_0^{\vmin^{\ast}} f_1(v) \~ dv}
        {\int_0^{\vmax} f_1(v) \~ dv}
\~.
\eeq
 20 different values of $\vmin^{\ast}$
 from 10 to 300 km/s
 for four most commonly used detector materials
 are given.

 As references,
 in Figs.~\ref{fig:f1v-Ge-025-0500-02-sh-sh_v0-flat-gen}
 to \ref{fig:f1v-Si-025-0500-05-sh-sh_v0-flat-gen}
 the reconstructed velocity distributions
 taking into account
 the corrections of the theoretical estimate of
 the difference between the triangular approximation
 and the integral over $f_1(v)$
 between 0 and $\vmin^{\ast}$,
 $\Delta_0^{\vmin^{\ast}} - \int_0^{\vmin^{\ast}} f_1(v) \~ dv$,
 have also been given as
 the dotted magenta horizontal lines.
 It can then be seen clearly that,
 with this (final) {\em model--dependent} correction
 for the normalization constant $\calN$,
 the reconstructed velocity distributions
 could match the true (input) distribution function
 very precisely:
 The tiny differences would totally be negligible,
 compared to the much larger statistical uncertainties
 given with ${\cal O}(500)$ WIMP events.

 Note however that,
 giving the correction of the systematic bias
 requires a theoretically predicted velocity distribution function.
 For practical use without prior knowledge
 about the one--dimensional WIMP velocity distribution
 as well as
 for improving the simple triangular approximation
 to the integral of the $v \le \vmin^{\ast}$ area
 (depending on the estimate of $f_1(v)$ at $v = \vmin^{\ast}$),
 one could use an iterative procedure with
 the Bayesian reconstructed velocity distribution function
 \cite{DMDDf1v-Bayesian}.
 Nevertheless,
 considering the pretty large statistical uncertainties
 on the reconstructed velocity distribution points
 as well as
 the much narrower statistical uncertainty band of
 the Bayesian reconstructed velocity distribution
 \cite{DMDDf1v-Bayesian},
 the effect of ignoring
 the much smaller systematic bias
 with or even without the model--dependent theoretical corrections
 could not be significant
 (at least in the next few years).

\begin{figure}[t!]
\begin{center}
\includegraphics[width=15cm]{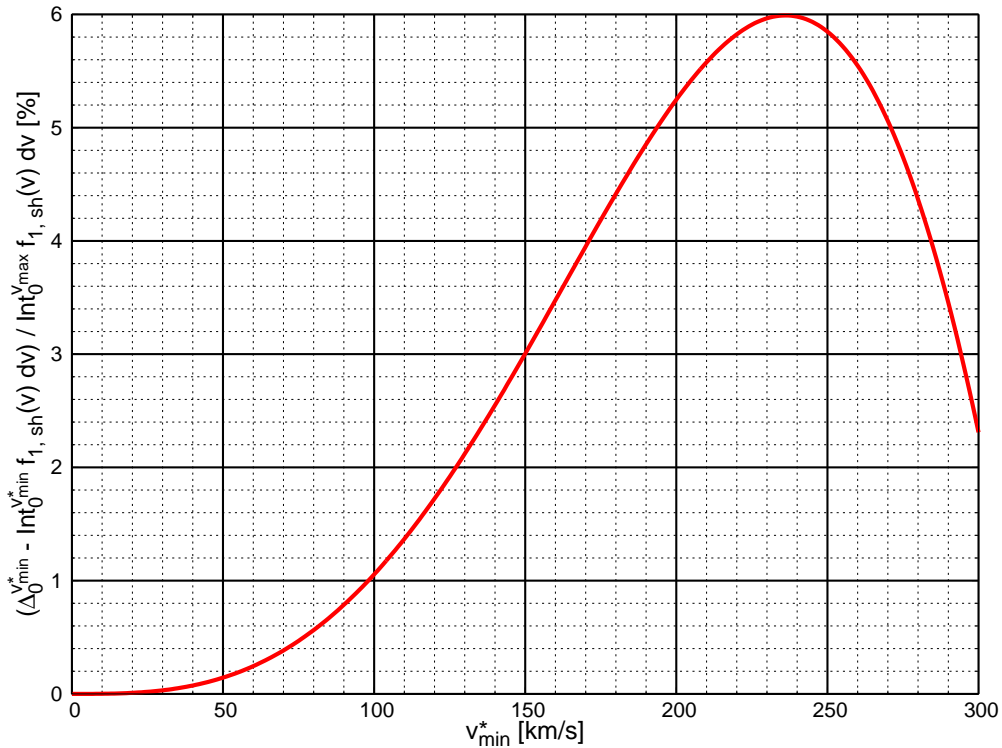} \\
\end{center}
\caption{
 The theoretically estimated fraction of
 the difference between
 the triangular approximation
 and the integral over $f_1(v)$
 between 0 and $\vmin^{\ast}$
 (i.e.~the little overestimated amount)
 to the integral in the entire velocity range
 between 0 and $\vmax$
 as a function of $\vmin^{\ast}$.
}
\label{fig:ratio_Delta}
\end{figure}

 Besides of Table 1,
 in Fig.~\ref{fig:ratio_Delta}
 we give the theoretically estimated fraction of
 the difference between
 the triangular approximation
 and the integral over $f_1(v)$
 between 0 and $\vmin^{\ast}$
 (i.e.~the little overestimated amount)
 to the integral in the entire velocity range
 between 0 and $\vmax$
 as a function of $\vmin^{\ast}$
 for reader's reference.

\section{Summary and conclusions}

 In this paper,
 we investigated the modification of our expressions
 developed for
 the model--independent data analysis procedure of
 the reconstruction of the (time--averaged) one--dimensional
 velocity distribution of Galactic
 WIMPs
 with a non--negligible experimental threshold energy.

 It our earlier work \cite{DMDDf1v},
 the experimental maximal and minimal cut--off energies
 have been assumed to be large or small enough.
 Thus
 the sum over all recorded events in the data set
 can be used as the estimator
 for the integral over the one--dimensional
 WIMP velocity distribution function,
 which is in turn needed for the estimation of
 the normalization constant of
 the reconstructed velocity distribution.
 For experiments with heavy target nuclei,
 e.g.~Ge or Xe,
 and once WIMPs are heavy
 (\mbox{$\gsim~100$ GeV}),
 the systematic bias caused by this assumption
 should be acceptable.
 However,
 once WIMPs are light
 (\mbox{$\lsim~50$ GeV})
 and a light target nucleus,
 e.g.~Si or Ar,
 is used
 for reconstructing the WIMP velocity distribution $f_1(v)$,
 effects of a non--negligible threshold energy
 has to be considered and
 the estimate of the normalization constant of $f_1(v)$
 would need in turn to be modified properly.

 In this work,
 we derived at first
 the expression for estimating the integral over $f_1(v)$
 between the minimal and maximal reconstructable velocities.
 Then
 we suggested the simple model--independent triangular approximation
 to the contribution below
 the minimal reconstructable velocity.
 Finally,
 by adopting the most commonly used
 shifted Maxiwellian velocity distribution function,
 the correction of the systematic bias
 caused by the use of the simple triangular approximation
 has been given.
 Our numerical simulations presented in this paper
 show that,
 for a minimal reconstructable velocity of
 as high as \mbox{${\cal O}(200)$ km/s},
 our model--independent modification of
 the estimator for the normalization constant
 (with or even without the model--dependent correction
  of the systematic bias)
 could provide
 precise reconstructed velocity distribution points
 to match the true WIMP velocity distribution
 with a \mbox{$\lsim$ 10\%} bias.

 In summary,
 as a supplement of our earlier works
 on the (Bayesian) reconstruction of
 the WIMP velocity distribution function,
 we developed in this paper
 a model--independent modification of
 the estimator for the normalization constant of $f_1(v)$
 for the more general case
 with a non--negligible experimental threshold energy.
 This modification
 should not only be more suitable
 for our Bayesian reconstruction of the one--dimensional
 WIMP velocity distribution function
 \cite{DMDDf1v-Bayesian},
 but hopefully also
 offer preciser information about Galactic Dark Matter
 for direct and indirect detection experiments and phenomenology.

\subsubsection*{Acknowledgments}
 The author would like to thank
 the Physikalisches Institut der Universit\"at T\"ubingen
 for the technical support of the computational work
 presented in this paper.
 This work
 was partially supported by
 the CAS Fellowship for Taiwan Youth Visiting Scholars
 under the grant no.~2013TW2JA0002
 as well as
 the Department of Human Resources and Social Security of
 Xinjiang Uygur Autonomous Region.

\appendix
\setcounter{equation}{0}
\setcounter{figure}{0}
\renewcommand{\theequation}{A\arabic{equation}}
\renewcommand{\thefigure}{A\arabic{figure}}
%
%
%
\section{Formulae for estimating statistical uncertainties}

 Here we list all formulae needed
 for the model--independent method
 for the reconstruction of
 the one--dimensional WIMP velocity distribution function
 described in Sec.~2
 as well as
 for the modification of the normalization constant $\calN$
 by Eq.~(\ref{eqn:calN_sum_mod})
 and the modified covariance matrix of
 the estimates of $f_{1, {\rm rec}}(v_{s, n})$.
 Detailed derivations and discussions
 can be found in Ref.~\cite{DMDDf1v}.

\subsection{Formulae needed in Sec.~2}

 First,
 by using the standard Gaussian error propagation,
 the expressions for the uncertainties on
 the standard estimator $r_n$
 and
 the logarithmic slope $k_n$ estimated by Eq.~(\ref{eqn:bQn})
 can be given
 directly as
\beq
   \sigma^2(r_n)
 = \frac{N_n}{b_n^2}
\~,
\label{eqn:sigma_rn}
\eeq
 and
\beq
   \sigma^2(k_n)
 = k_n^4
   \cbrac{  1
          - \bfrac{k_n b_n / 2}{\sinh (k_n b_n / 2)}^2}^{-2}
            \sigma^2\abrac{\bQn}
\~,
\label{eqn:sigma_kn}
\eeq
 where
\beq
   \sigma^2\abrac{\bQn}
 = \frac{1}{N_n - 1} \bbigg{\bQQn - \bQn^2}
\~.
\label{eqn:sigma_bQn}
\eeq
 For replacing the ``bin'' quantities
 by ``window'' quantities,
 one needs the covariance matrix
 for $\Bar{Q - Q_{\mu}}|_{\mu}$,
 which follows directly from the definition (\ref{eqn:wQ_mu}):
\beqn
 \conti {\rm cov}\abrac{\Bar{Q - Q_{\mu}}|_{\mu}, \Bar{Q - Q_{\nu}}|_{\nu}}
        \non\\
 \=     \frac{1}{N_{\mu} N_{\nu}}
        \sum_{n = n_{\nu-}}^{n_{\mu+}}
        \bbigg{  N_n
                 \abrac{\Bar{Q}|_n - \Bar{Q}|_{\mu}}
                 \abrac{\Bar{Q}|_n - \Bar{Q}|_{\nu}}
               + N_n^2 \sigma^2\abrac{\bQn}}
\~.
\label{eqn:cov_wQ_mu}
\eeqn
 Note that,
 firstly,
 $\mu \leq \nu$ has been assumed here
 and the covariance matrix is, of course, symmetric.
 Secondly,
 the sum is understood to vanish
 if the two windows $\mu$, $\nu$ do not overlap,
 i.e.~if $n_{\mu+} < n_{\nu-}$.
 Moreover,
 similar to Eq.~(\ref{eqn:sigma_rn}),
 we can get
\beq
   {\rm cov}(r_{\mu}, r_{\nu})
 = \frac{1}{w_{\mu} w_{\nu}} \sum_{n = n_{\nu-}}^{n_{\mu+}} N_n
\~,
\label{eqn:cov_r_mu}
\eeq
 where $\mu \leq \nu$ has again been taken.
 And the mixed covariance matrix can be given by
\beq
   {\rm cov}\abrac{r_{\mu}, \Bar{Q - Q_{\nu}}|_{\nu}}
 = \frac{1}{w_{\mu} N_{\nu}}
   \sum_{n = n_{-}}^{n_{+}} N_n \abrac{\Bar{Q}|_n - \Bar{Q}|_{\nu}}
\~.
\label{eqn:cov_r_mu_wQ_nu}
\eeq
 Note here that
 this sub--matrix is {\em not} symmetric
 under the exchange of $\mu$ and $\nu$.
 In the definition of $n_{-}$ and $n_{+}$
 we therefore have to distinguish two cases:
\beqn
\renewcommand{\arraystretch}{1.6}
\begin{array}{l c l}
   n_{-} = n_{\nu-},~
   n_{+} = n_{\mu+},  &
   ~~~~~~~~~~~~       &
   {\rm if}~\mu \leq \nu\~; \\
   n_{-} = n_{\mu-},~
   n_{+} = n_{\nu+},  &
                      &
   {\rm if}~\mu \geq \nu\~.
\end{array}
\label{eqn:def_n_pm}
\eeqn
 As before,
 the sum in Eq.~(\ref{eqn:cov_r_mu_wQ_nu})
 is understood to vanish if $n_{-} > n_{+}$.

 Furthermore,
 the covariance matrices
 involving the estimators of the logarithmic slopes $k_{\mu}$,
 estimated by Eq.~(\ref{eqn:bQn}) with replacing $n \to \mu$,
 can be given from Eq.~(\ref{eqn:sigma_kn}) as
\beqn
        {\rm cov}\abrac{k_{\mu}, k_{\nu}}
 \=     k_{\mu}^2 k_{\nu}^2
        \cbrac{  1
               - \bfrac{k_{\mu} b_{\mu} / 2}{\sinh (k_{\mu} b_{\mu} / 2)}^2}^{-1}
        \cbrac{  1
               - \bfrac{k_{\nu} b_{\nu} / 2}{\sinh (k_{\nu} b_{\nu} / 2)}^2}^{-1}
        \non\\
        \non\\
 \conti ~~~~~~~~~~~~~~~~ \times 
        {\rm cov}\abrac{\Bar{Q - Q_{\mu}}|_{\mu}, \Bar{Q - Q_{\nu}}|_{\nu}}
\~,
\label{eqn:cov_k_mu}
\eeqn
 and
\beq
   {\rm cov}\abrac{r_{\mu}, k_{\nu}}
 = k_{\nu}^2
   \cbrac{  1
          - \bfrac{k_{\nu} b_{\nu} / 2}{\sinh (k_{\nu} b_{\nu} / 2)}^2}^{-1}
   {\rm cov}\abrac{r_{\mu}, \Bar{Q - Q_{\nu}}|_{\nu}}
\~.
\label{eqn:cov_r_mu_k_nu}
\eeq
\subsection{Derivatives of the modified estimates
            \boldmath$f_{1, {\rm rec}}(v_{s, \mu})$}

 The modified normalization constant $\calN$
 given by Eq.~(\ref{eqn:calN_sum_mod})
 depends on the estimates of $r_1$ and $k_1$,
 as the first reconstructed point of the velocity distribution
 given in Eq.~(\ref{eqn:f1v_Qsn}),
 $f_{1, {\rm rec}}(v_{s, 1})$.
 For modifying the covariance matrix of the estimates of $f_1(v)$,
 one needs thus to distinguish the $\mu = 1$ case
 from the other $\mu \neq 1$ cases.

 First,
 for the general $\mu \neq 1$ case,
 one has
\beqn
    f_{1, {\rm rec}}(v_{s, \mu})
 \=  \frac{2}{\alpha}
     \bbrac{  \tilde{f}_{1, {\rm rec}}(\vmin^{\ast}) \~ \Qmin^{1 / 2}
            + \frac{2 \Qmin^{1 / 2} r(\Qmin)}{\FQmin}
            + I_0(\Qmin, \Qmax^{\ast})}^{-1}
     \non\\
 \conti ~~~~~~~~~~~~ \times 
     \bBigg{\frac{2 Q_{s, \mu} r_{\mu}}{F^2(Q_{s, \mu})}}
     \bbrac{\dd{Q} \ln \FQ \bigg|_{Q = Q_{s, \mu}} - k_{\mu}}
     \non\\
 \=  \frac{2}{\alpha}
     \cleft{   \bBigg{\frac{2 \Qmin^{1 / 2} r_1 e^{k_1 (\Qmin - Q_{s, 1})}}{F^2(\Qmin)}}
               \cbrac{  \bbrac{\dd{Q} \ln \FQ \bigg|_{Q = \Qmin} - k_1}
                        \Qmin
                      + 1} }
     \non\\
 \conti ~~~~~~~~~~~~~~~~~~~~~~~~ 
     \cBiggr{+ I_0(\Qmin, \Qmax^{\ast})}^{-1}
     \non\\
 \conti ~~~~~~~~~~~~ \times 
     \bBigg{\frac{2 Q_{s, \mu} r_{\mu}}{F^2(Q_{s, \mu})}}
     \bbrac{\dd{Q} \ln \FQ \bigg|_{Q = Q_{s, \mu}} - k_{\mu}}
\~.
\label{eqn:f1v_Qs_mu_mod}
\eeqn
 Then
 the derivative of $f_{1, {\rm rec}}(v_{s, \mu})$
 to $r_1$ can be given as
\cheqnXa{A}
\beqn
     \Pp{f_{1, {\rm rec}}(v_{s, \mu})}{r_1}
 \=- \frac{2}{\alpha}
     \cbigg{\cdots}^{-2}
     \bBigg{\frac{2 \Qmin^{1 / 2} e^{k_1 (\Qmin - Q_{s, 1})}}{F^2(\Qmin)}}
     \cbrac{  \bbrac{\dd{Q} \ln \FQ \bigg|_{Q = \Qmin} - k_1}
              \Qmin
            + 1}
     \non\\
 \conti ~~~~~~~~~~~~ \times 
     \bBigg{\frac{2 Q_{s, \mu} r_{\mu}}{F^2(Q_{s, \mu})}}
     \bbrac{\dd{Q} \ln \FQ \bigg|_{Q = Q_{s, \mu}} - k_{\mu}}
     \non\\
 \=- \frac{f_{1, {\rm rec}}(v_{s, \mu})}{r_1}
     \bbrac{1 - \calN \afrac{\alpha}{2} I_0(\Qmin, \Qmax^{\ast})}
\~.
\label{eqn:df1v_Qs_mu_mod_dr_1}
\eeqn
 And
 the derivative of $f_{1, {\rm rec}}(v_{s, \mu})$ to $k_1$ is
\cheqnXb{A}
\beqn
     \Pp{f_{1, {\rm rec}}(v_{s, \mu})}{k_1}
 \=- \frac{2}{\alpha}
     \cbigg{\cdots}^{-2}
     \bBigg{\frac{2 \Qmin^{1 / 2} r_1 e^{k_1 (\Qmin - Q_{s, 1})}}{F^2(\Qmin)}}
     \non\\
 \conti ~~~~~~ \times 
     \cbrac{  \abrac{\Qmin - Q_{s, 1}}
              \cbrac{  \bbrac{\dd{Q} \ln \FQ \bigg|_{Q = \Qmin} - k_1}
                       \Qmin
                     + 1}
            - \Qmin}
     \non\\
 \conti ~~~~~~~~~~~~ \times 
     \bBigg{\frac{2 Q_{s, \mu} r_{\mu}}{F^2(Q_{s, \mu})}}
     \bbrac{\dd{Q} \ln \FQ \bigg|_{Q = Q_{s, \mu}} - k_{\mu}}
     \non\\
 \=- f_{1, {\rm rec}}(v_{s, \mu})
     \bbrac{\calN \afrac{\alpha}{2}}
     \bBigg{\frac{2 \Qmin^{1 / 2} r(\Qmin)}{F^2(\Qmin)}}
     \non\\
 \conti ~~~~~~ \times 
     \cbrac{  \abrac{\Qmin - Q_{s, 1}}
              \bbrac{\dd{Q} \ln \FQ \bigg|_{Q = \Qmin} - k_1}
              \Qmin
            - Q_{s, 1}}
\~.
\label{eqn:df1v_Qs_mu_mod_dk_1}
\eeqn
\cheqnX{A}%
 Note that
 once $\Qmin = 0$,
 the second term in the bracket
 in the second line of Eq.~(\ref{eqn:df1v_Qs_mu_mod_dr_1})
 reduces to 1
 and $\p f_{1, {\rm rec}}(v_{s, \mu}) / \p r_1$
 as well as
 $\p f_{1, {\rm rec}}(v_{s, \mu}) / \p k_1$
 become 0.
 Moreover,
 similar to
 the calculations done for the covariance matrix
 in Eq.~(\ref{eqn:cov_f1v_Qs_mu}),
 one can get
\cheqnXa{A}
\beq
    \Pp{f_{1, {\rm rec}}(v_{s, \mu})}{r_{\mu}}
 =  \frac{f_{1, {\rm rec}}(v_{s, \mu})}{r_{\mu}}
\~,
\label{eqn:df1v_Qs_mu_mod_dr_mu}
\eeq
 and
\cheqnXb{A}
\beq
    \Pp{f_{1, {\rm rec}}(v_{s, \mu})}{k_{\mu}}
 =- \calN \bBigg{\frac{2 Q_{s, \mu} r_{\mu}}{F^2(Q_{s, \mu})}}
\~.
\label{eqn:df1v_Qs_mu_mod_dk_mu}
\eeq
\cheqnX{A}
 On the other hand,
 for the special $\mu = 1$ case,
 we have
\beqn
    f_{1, {\rm rec}}(v_{s, 1})
 \=  \frac{2}{\alpha}
     \cleft{   \bBigg{\frac{2 \Qmin^{1 / 2} r_1 e^{k_1 (\Qmin - Q_{s, 1})}}{F^2(\Qmin)}}
               \cbrac{  \bbrac{\dd{Q} \ln \FQ \bigg|_{Q = \Qmin} - k_1}
                        \Qmin
                      + 1} }
     \non\\
 \conti ~~~~~~~~~~~~~~~~~~~~~~~~ 
     \cBiggr{+ I_0(\Qmin, \Qmax^{\ast})}^{-1}
     \non\\
 \conti ~~~~~~~~~~~~ \times 
     \bBigg{\frac{2 Q_{s, 1} r_1}{F^2(Q_{s, 1})}}
     \bbrac{\dd{Q} \ln \FQ \bigg|_{Q = Q_{s, 1}} - k_1}
\~.
\label{eqn:f1v_Qs_1_mod}
\eeqn
 Then,
 similar to
 the calculations done
 in Eqs.~(\ref{eqn:df1v_Qs_mu_mod_dr_1})
 to (\ref{eqn:df1v_Qs_mu_mod_dk_mu}),
 it can be found that
\cheqnXa{A}
\beq
    \Pp{f_{1, {\rm rec}}(v_{s, 1})}{r_1}
 =  \left. \Pp{f_{1, {\rm rec}}(v_{s, \mu})}{r_1}   \right|_{\mu = 1}
  + \left. \Pp{f_{1, {\rm rec}}(v_{s, \mu})}{r_{\mu}} \right|_{\mu = 1}
\~,
\label{eqn:df1v_Qs_1_mod_dr_1}
\eeq
 and
\cheqnXb{A}
\beq
    \Pp{f_{1, {\rm rec}}(v_{s, 1})}{k_1}
 =  \left. \Pp{f_{1, {\rm rec}}(v_{s, \mu})}{k_1}   \right|_{\mu = 1}
  + \left. \Pp{f_{1, {\rm rec}}(v_{s, \mu})}{k_{\mu}} \right|_{\mu = 1}
\~.
\label{eqn:df1v_Qs_1_mod_dk_1}
\eeq
\cheqnX{A}

\begin{thebibliography}{99}
%
\bibitem{SUSYDM96}
 {G.~Jungman, M.~Kamionkowski and K.~Griest,
  {\it ``Supersymmetric Dark Matter''},
  {\it Phys.~Rep.}~{\bf 267}, 195--373 (1996),
  {\tt arXiv:hep-ph/9506380}.}
%
\bibitem{Drees12}
 {M.~Drees and G.~Gerbier,
  {\it ``Mini--Review of Dark Matter: 2012''},
  updated minireview for
  {\it ``The Review of Particle Physics 2012''},
  {\tt arXiv:1204.2373 [hep-ph]} (2012).}
%
\bibitem{Strigari12b}
 {L.~E.~Strigari,
  {\it ``Galactic Searches for Dark Matter''},
  {\it Phys.~Rep.}~{\bf 531}, 1--88 (2013),
  {\tt arXiv:1211.7090 [astro-ph.CO]}.}
%
\bibitem{Baudis12c}
 {L.~Baudis,
  {\it ``Direct Dark Matter Detection: the Next Decade''},
  Issue on {\it ``The Next Decade in Dark Matter and Dark Energy''},
  {\it Phys.~Dark Univ.}~{\bf 1}, 94--108 (2012),
  {\tt arXiv:1211.7222 [astro-ph.IM]}.}
%
%
\bibitem{DMDDf1v}
 {M.~Drees and C.-L.~Shan,
  {\it ``Reconstructing the Velocity Distribution of Weakly Interacting Massive Particles
         from Direct Dark Matter Detection Data''},
  {\it J.~Cosmol.~Astropart.~Phys.}~{\bf 0706}, 011 (2007),
  {\tt arXiv:astro-ph/0703651}.}
%
\bibitem{DMDDf1v-Bayesian}
 {C.-L.~Shan,
  {\it ``Bayesian Reconstruction of the Velocity Distribution of
         Weakly Interacting Massive Particles
         from Direct Dark Matter Detection Data''},
  {\it J.~Cosmol.~Astropart.~Phys.}~{\bf 1408}, 009 (2014),
  {\tt arXiv:1403.5610 [astro-ph.HE]}.}
%
%
\bibitem{DMDDmchi}
 {M.~Drees and C.-L.~Shan,
  {\it ``Model--Independent Determination of the WIMP Mass
         from Direct Dark Matter Detection Data''},
  {\it J.~Cosmol.~Astropart.~Phys.}~{\bf 0806}, 012 (2008),
  {\tt arXiv:0803.4477 [hep-ph]}.}
%
%
\bibitem{Freese88}
 {K.~Freese, J.~Frieman and A.~Gould,
  {\it ``Signal Modulation in Cold--Dark--Matter Detection''},
  {\it Phys.~Rev.}~{\bf D37}, 3388--3405 (1988).}
%
%
\bibitem{AMIDAS-web}
 {C.-L.~Shan,
  the {\tt AMIDAS} (A Model--Independent Data Analysis System)
  package and website for direct Dark Matter detection experiments and phenomenology, \\
  {\tt http://pisrv0.pit.physik.uni-tuebingen.de/darkmatter/amidas/} (2009); \\ 
%
  the mirror website on TiResearch (Taiwan interactive Research), \\
  {\tt http://www.tir.tw/phys/hep/dm/amidas/}.}
%
%
\bibitem{AMIDAS-II}
 {C.-L.~Shan,
  {\it ``AMIDAS-II:
         Upgrade of the AMIDAS Package and Website
         for Direct Dark Matter Detection Experiments and Phenomenology''},
  {\it Phys.~Dark Univ.}~{\bf 5--6}, 240--306 (2014),
  {\tt arXiv:1403.5611 [astro-ph.IM]}.}
%
%
\bibitem{Lisanti10}
 {M.~Lisanti, L.~E.~Strigari, J.~G.~Wacker and R.~H.~Wechsler,
  {\it ``The Dark Matter at the End of the Galaxy''},
  {\it Phys.~Rev.}~{\bf D 83}, 023519 (2011),
  {\tt arXiv:1010.4300 [astro-ph.CO]}.}
%
%
\bibitem{YYMao}
 {Y.-Y.~Mao, L.~E.~Strigari, R.~H.~Wechsler, H.-Y.~Wu and O.~Hahn,
  {\it ``Halo--to--Halo Similarity and Scatter in the Velocity Distribution of Dark Matter''}
  {\it Astrophys.~J.}~{\bf 764}, 35 (2013),
  {\tt arXiv:1210.2721 [astro-ph.CO]};} \\
%
 {Y.-Y.~Mao, L.~E.~Strigari and R.~H.~Wechsler,
  {\it ``Connecting Direct Dark Matter Detection Experiments to
         Cosmologically Motivated Halo Models''},
  {\it Phys.~Rev.}~{\bf D 89}, 063513 (2014),
  {\tt arXiv:1304.6401 [astro-ph.CO]}.}
%
\bibitem{Kuhlen13}
 {M.~Kuhlen, A.~Pillepich, J.~Guedes and P.~Madau,
  {\it ``The Distribution of Dark Matter in the Milky Way's Disk''},
  {\it Astrophys.~J.}~{\bf 784}, 161 (2014),
  {\tt arXiv:1308.1703 [astro-ph.GA]}.}
%
%
\end{thebibliography}
\end{document}